\documentclass[prb,twocolumn,superscriptaddress,preprintnumbers,amsmath,amssymb]{revtex4}
\usepackage{color}
\usepackage{graphicx}
\usepackage{amsmath}
\usepackage{amssymb}
\usepackage{epsfig}
\usepackage{epstopdf}
\usepackage{wasysym}
\usepackage{bbm}
\usepackage{multirow}
\renewcommand{\narrowtext}{\begin{multicols}{2} \global\columnwidth20.5pc}

\newcommand{\s}{{\sigma}}

\def\be{\begin{eqnarray}}
\def\ee{\end{eqnarray}}

\newcommand{\nn}{\nonumber\\}

\newcommand{\Eq}[1]{Eq.~(\ref{#1})}

\newcommand{\ra}{\rightarrow}

\newcommand{\Fig}[1]{Fig.~\ref{#1}}
\newcommand{\Figs}[1]{Figs.~\ref{#1}}

\begin{document}

\title{The Quantum Torus Chain}
\author{M.P. Qin}
\affiliation{%
Institute of Physics, Chinese Academy of Sciences, P.O. Box 603, Beijing 100190, China
}
\author{J.M. Leinaas}
\affiliation{%
Department of Physics, University of Oslo, P.O. Box 1048 Blindern, N-0316 Oslo, Norway
}

\author{S. Ryu}
\affiliation{%
Department of Physics, University of Illinois at Urbana-Champaign,
1110 W. Green Street, Urbana, Illinois 61801-3080, USA
}
\author{E. Ardonne}
\affiliation{%
Nordita, Royal Institute of Technology and Stockholm University, Roslagstullsbacken 23, SE-10691 Stockholm, Sweden
}
\affiliation{%
Department of Physics, Stockholm University, AlbaNova University Center, SE-106 91 Stockholm, Sweden
}

\author{T. Xiang}
\affiliation{%
Institute of Physics, Chinese Academy of Sciences, P.O. Box 603, Beijing 100190, China
}
\affiliation{%
Institute of Theoretical Physics, Chinese Academy of Sciences, P.O. Box 2735, Beijing 100190, China
}
\author{D.-H. Lee}
\affiliation{%
Department of Physics, University of California
at Berkeley, Berkeley, California 94720, USA
}
\affiliation{%
Materials Sciences Division, Lawrence Berkeley
National Laboratory, Berkeley, California 94720, USA
}

\begin{abstract}
We introduce a new set of one dimensional quantum lattice models which we refer to as
{\em The quantum torus chain}. These models have discrete global symmetry, and {\em projective} on-site representations. They possess an integer-valued parameter which controls the presence or absence of frustration. Depending on whether this parameter is even or odd these models either exhibit gapped symmetry breaking phases with isolated critical points, or gapped symmetry breaking phases separated by gapless {\it phases}. We discuss the property of these phases and phase transitions for two special values of the parameter and point out many open problems.

\end{abstract}

\date{\today}
\maketitle

\section{Introduction}
Over the years many  interesting lattice models have been introduced to capture the essence of important physical concepts, and make them open for more quantitative studies. In modern language quantum lattice models can capture short-range entangled states such as symmetry breaking or symmetry protected topological states\cite{Chen11}. It also can capture long-range entangled states such as quantum critical and the ``intrinsic topological ordered'' states.

The importance of symmetry is well known in classical and quantum physics. Recently it is realized that a strong tie exists between quantum entanglement and symmetry.\cite{Chen11} For example, in one space dimension Chen {\it et al}\cite{Chen11} showed that short-range entangled, fully symmetric, quantum states are classified by the {\it projective} representation of the internal symmetry group. In addition, if the site representation of the symmetry group  is {\it projective}, short-range entanglement is impossible without symmetry breaking. Because in one dimension a gapped system is necessarily short-range entangled, this implies {{\em projective}} on-site representation can not have an energy gap without breaking some symmetry.  This result is a generalization of Haldane's work on SO(3) spin chain\cite{hal}. There half integer spin chains need to break a symmetry to open an energy gap, while integer spin chains do not.

In this paper we introduce a new family of one dimensional quantum lattice models whose global symmetry groups
are discrete and the on-site representation is  projective. Indeed, these models are either gapless or exhibit spontaneously symmetry breaking. In addition, they possess an integer-valued parameter which controls the presence or absence of frustration. Depending on whether this parameter is even or odd, qualitatively different phase diagrams are observed. By applying the density matrix renormalization group (DMRG)\cite{DMRG_1,DMRG_2} and Matrix product state (MPS)\cite{MPS} methods, we find for even parameter these models are generically gapped and show spontaneous symmetry breaking. Fine tuning is required to close the energy gap. On the other hand for odd parameter, both gapped symmetry breaking phase and gapless symmetric phase are generic.
The results are all consistent with the conclusions drawn by Chen {\it et. al} \cite{Chen11}.

The results reported in this paper show that the new class of models that we introduced has a rich set of interesting phases. We hope that the study of this new class of models will enhance our understanding of symmetry protected topological phases in one dimensional systems, and of topological phases in a more general setting.

\section{The quantum torus chain}

The family of models, which we refer to as the quantum torus chain,
is motivated from the following view of the quantum spin chain.

The dimension $D$ of the single site Hilbert space of a
$\mathrm{SU}(2)$ spin-$S$ chain is $2S+1$. The basis states can be viewed as the one-particle states of a unit-charge particle running on a sphere enclosing a magnetic monopole which produces $2S$ magnetic flux quanta through the sphere\cite{YW}. The nonzero total Gaussian curvature of the sphere leads to a term in the Hamiltonian which has the same form as the term describing the flux. Thus, the curvature effectively increases the flux, and there are $D=2S+1$ rather than $2S$ one-particle states (see also \cite{AS}).
The spin operators $S^{x,y,z}$ are the generators of the magnetic translations.

\begin{figure}[tbp]
\begin{center}
\vskip-0.5in
\includegraphics[height=2in]{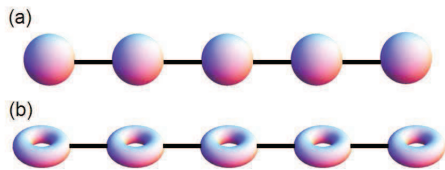}
\vskip-0.5in\caption{(a) Each site of a quantum spin-$S$ chain can be viewed as the lowest Landau level of a unit charged particle moving on a sphere under the magnetic field of a Dirac monopole.
The total magnetic flux produced by the monopole is $2S$. (b) The Torus chain is a
modification of the spin chain where each sphere is replaced by a torus. \label{ft}}
\end{center}
\end{figure}

In the XYZ model the nearest neighboring spin operators are coupled as
\be H_{XYZ}=\sum_i (J_x S^x_iS^x_{i+1}+J_y S^y_iS^y_{i+1}+J_z S^z_iS^z_{i+1}).\ee
This is depicted in \Fig{ft}(a).

In this work we consider the ``torus chain'' (\Fig{ft}(b)),
where the spheres of \Fig{ft}(a) are replaced by the torus.
Because the total Gauss curvature of a torus is zero, the lowest Landau level degeneracy is equal to the number of
flux quanta through
each torus\cite{AS}.
Unlike the sphere, the magnetic translation group of a torus, the Heisenberg-Weyl group, is {\it discrete} rather than continuous\cite{ref}. The commutation relation between the operators $U^x,U^y$ which generate the magnetic translation in two orthogonal directions is
\begin{eqnarray}
U^xU^y=e^{i2\pi/m}U^yU^x,
\end{eqnarray}
where $(U^x)^m=(U^y)^m=I$.
The $m$ in the preceding equation is the number of
flux quanta through
each torus, and is the integer valued parameter we referred to earlier.
In the following we shall encounter two different bases, $|q\rangle$ and $|\tilde{q}\rangle$, for the site-Hilbert space. Both $q$ and $\tilde{q}$ are defined modulo $m$.
In the $|q\rangle$ basis, $U^x$  and $U^y$ are given by
\be
&&U^x=\sum_{q=0}^{m-1}e^{i2\pi q/m}|q\rangle\langle q|,
\nn
&&U^y=\sum_{q=0}^{m-1} |q+1\rangle\langle q|.\ee
The dual basis $|\tilde{q}\rangle$ is the Fourier transform of $|q\rangle$,
i.e.
$$
|\tilde{q}\rangle = \frac{1}{\sqrt{m}} \sum_{q=0}^{m-1} e^{2 \pi i q \tilde{q}/m} |q\rangle \ .
$$
In this basis
\be
&&
U^x=\sum_{\tilde{q}=0}^{m-1} |\tilde q+1\rangle\langle \tilde{q}|,
\nn
&&
U^y = \sum_{\tilde{q}=0}^{m-1} e^{-2\pi i \tilde{q}/m} |\tilde{q} \rangle\langle\tilde{q}|.
\ee

Analogously to
the XYZ model
we couple the generators of the magnetic translation group of the torus to form the following Hamiltonian:
 \be
H=\sum_i (\cos\theta ~U^x_i U^{x \dag}_{i+1}+\sin\theta ~U^y_i U^{y\dag}_{i+1}+h.c.).
\label{tr}
\ee
Here,
the parameter $\theta$ is introduced to control the
relative strength (and sign)
between two terms,
$\sum_i U^x_i U^{x \dag}_{i+1}+ h.c.$
and
$\sum_i U^y_i U^{y\dag}_{i+1}+h.c.$,
in the Hamiltonian.
In the following we assume the number of sites $L$ to be an integer multiple of $m$. With this restriction the two unitary operators
${\cal U}^x=\prod_{i=1}^L U^x_i$
and
${\cal U}^y=\prod_{i=1}^L U^y_i$ commute with each other and with the Hamiltonian (\ref{tr}).
The Hamiltonian (\ref{tr}) is one of the simplest models
which preserve the full group of magnetic translations
(the group generated by ${\cal U}^x$ and ${\cal U}^y$),
{\it i.e.} an analogue of an $\mathrm{SU}(2)$ symmetric spin chain.
It is possible to consider a further generalization of the model, with the same symmetry, by
adding different complex phase factors
for the two terms. However for simplicity we shall restrict ourselves to the case where the coupling ratio is real. We also note the similarity of our model to
the  $m$-state Potts model
(although the symmetries  are a bit different).
For more descriptions about related/similar models, see, {\it e.g.},
Refs.\ \onlinecite{resh, su3_1, Ortiz2011,Mansson2011}.

We will denote the conserved quantum numbers associated with $\mathcal{U}^x$ and
$\mathcal{U}^y$ by $(e^{2\pi i Q/m},e^{-2\pi i \tilde{Q}/m})$, or more compactly as
$(Q,\tilde{Q})$. In terms of $q_i$ and $\tilde{q}_i$ we have
$$
Q = \sum_{i} q_i \bmod m \ , \quad \tilde{Q} = \sum_{i} \tilde{q}_i \bmod m \ .
$$
In addition to ${\cal U}^x$ and ${\cal U}^y$ there are
other symmetry operators that leave
the Hamiltonian (\ref{tr}) invariant.
First of all, there is the $q$-inversion,
${\cal R}=\prod_{i=1}^L R_i$ , where $$R_i|{q}_i\rangle=|-{q}_i \rangle {\rm ~mod~}m,~
R_i |\tilde{q}_i\rangle=|{-\tilde q}_i\rangle {\rm~mod~}m.$$
Secondly,
there are the following anti-unitary operations,
$K$ and $\tilde{K}=RK$, which cause complex conjugation in the
$q$ and $\tilde q$-basis respectively. We note that $K^2=\tilde K^2=I$.
The operators ${\cal U}^x$, ${\cal U}^y$, ${\cal R}$, and $K$
generate the internal symmetry group $G$.
As one can readily check the group multiplications of $G$ are identical to those of the symmetry group of a rectangular, periodic lattice on a torus, with ${\cal U}^x$ and ${\cal U}^y$ as the lattice translations, $R$ as rotation of angle $\pi$, and $K$ and $\tilde K$ as the two reflections. We note, as usual, the translation generators of the space group commute. This is because we have restricted $L$ to integer multiple of $m$. It is also interesting to note, unlike the usual cases, some of the space group elements are represented anti-unitarily here.
The internal symmetry group $G$ extended with the translations and inversion of the one-dimensional (1D) chain
defines the full symmetry group of the Hamiltonian.
In addition the Hamiltonian (\ref{tr}) has a ``duality'' symmetry upon $\theta\leftrightarrow \pi/2-\theta\;(mod\;2\pi)$ and $U^x\leftrightarrow U^y$.

The irreducible representations of the group of internal symmetries determine the (minimal)
degeneracies of the energy levels. For example for $m=3$, which we shall spend great length discussing, the multiplets take the following form in terms of the $(Q,\tilde{Q})$
\begin{align*}
(Q,\tilde{Q}) = & (0,0) \\
(Q,\tilde{Q}) = & (1,0) ; (2,0) \\
(Q,\tilde{Q}) = & (0,1) ; (0,2) \\
(Q,\tilde{Q}) = & (1,1) ; (1,2) ; (2,1) ; (2,2) \\
\end{align*}
At the ``self-dual'' point $\tan\theta = 1$ the two terms in the
Hamiltonian (\ref{tr}) have the same coefficient. There  the
symmetry group is larger, which causes the two doublets in the above table to become degenerate, and form
a quadruplet.

\section{Some properties for general $m$}

Having introduced the model and its symmetries, we start our analysis by
first considering the dependence of the model on the parameter $m$. It turns out that even and odd $m$ have qualitatively different phase diagrams.

For later discussions it is useful to consider the subgroup ${\cal G}$
generated by the subset of generators that are unitary:
$\{I,{\cal U}^x,{\cal U}^y,{\cal R}\}$.
This group is isomorphic to the symmetry group of an $m\times m$ oblique lattice on a torus.
It is important to note, however, for each site of the 1D chain this symmetry group is represented by a {\em projective} representation.
The non-trivial $U(1)$ phases are due to the presence of the factor $e^{i2\pi/m}$ in $U^xU^y=e^{i2\pi/m}U^yU^x$.
Applying the results of Chen {\it et al}
we {{therefore} conclude that \Eq{tr} should either exhibit a gapped spectrum with a symmetry breaking ground state, or a gapless spectrum with a symmetric ground state.

To see how this is realized, we  first consider the simplest case, with $m=2$. In this case we can rewrite \Eq{tr} in terms of a spin $1/2$ Hamiltonian. This is achieved by identifying $q=0,1$ with spin up and down, and express $U^{x,y}$ in terms of
the Pauli matrices $U^x\ra \s^z, U^y\ra\s^x.$
Under this identification \Eq{tr} becomes
\be
H\ra 2\cos\theta \sum_i \s^z_i \s^z_{i+1} +2\sin\theta \sum_i\s^x_i\s^x_{i+1}.\nonumber\ee
This is the anisotropic XZ model, which is gapped for all $\theta$ except $\theta=\pm\pi/4$ and $\pm 5\pi/4$ where it is quantum critical.
 The phase diagram, which is symmetric under $\theta\rightarrow \pi/2-\theta$ due to duality,  is shown in \Fig{phase}(a).
 \begin{figure}[tbp]
\begin{center}
\includegraphics[width=\columnwidth]{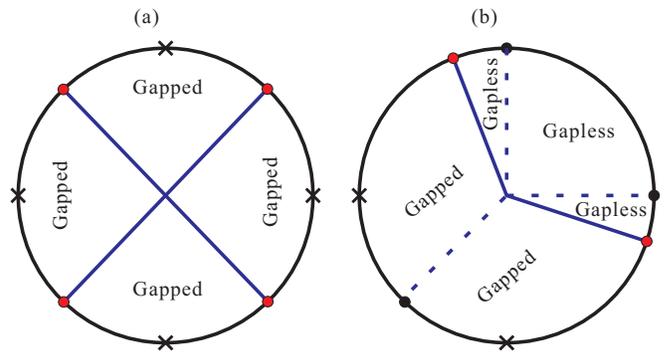}
\caption{ The phase diagrams of \Eq{tr} for (a) $m=2$ and (b) $m=3$.
The black and red dots mark first order transitions and critical points, respectively.
At points marked by crosses the model is classical hence exactly solvable.
\label{phase}}
\end{center}
\end{figure}
In the gapped phases the ${\cal U}^x$/${\cal U}^y$ and/or the translation symmetry is spontaneously broken leading to two-fold ground state degeneracy.  At the points $\theta=\pm\pi/4$ and $\theta = \pm5\pi/4$ the ground state respects all symmetries but the energy spectrum is gapless.

To get a qualitative understanding of  how this phase diagram generalizes to other values of $m$, it is instructive to consider four special points. First, for $\theta=\pi$ the Hamiltonian becomes classical because there are no non-commuting operators. In that case the ground state is given by $\prod_{i=1}^L |q\rangle$ and is $m$-fold degenerate for any given $m$. Clearly the ${\cal U}^y$ symmetry is spontaneously broken and there is an energy gap.
The gap persists for small $\theta-\pi$, which shows that this is an extended gapped phase. For $m=2$ it extends to the full interval $3\pi/4<\theta<5\pi/4$, with an equivalent, dual phase in the interval $5\pi/4<\theta<7\pi/4$.

Provided  $m$ is even the situation is similar at the point $\theta=0$. In that case  the ground state is also $m$-fold degenerate, now with a ground state of the form $\prod_{i=1}^L |(-1)^i \,q \rangle$. Also around the point $\theta=0$ there is a gapped phase, with a ground state that, in the $L\to\infty$ limit, spontaneously breaks both ${\cal U}^y$ and translational symmetry. For $m=2$ this phase also extends to the full interval $-\pi/4<\theta<\pi/4$, with the equivalent, dual phase in the interval $\pi/4<\theta<3\pi/4$

However, if $m$ is odd the situation is quite different.
For $\theta=0$, all states with  $$q_i-q_{i+1}= (m\pm 1)/2~({\rm mod}\; m)$$ yield the same energy. As the result, the ground state is extensively degenerate. For example, in an open chain the degeneracy is $2^{L-1}\,m$.  When $\theta$ deviates from zero these degenerate configurations are mixed by the $\sin\theta$ term, and hence the degeneracy is lifted. This is very similar to the effects of quantum fluctuations in frustrated magnets. Due to the duality of the Hamiltonian, the same discussions hold for $\theta=\pi/2$, except the roles of $q$ and $\tilde{q}$ are interchanged.

\section{The case $m=3$}

In the previous section, we used symmetry considerations to gain knowledge about the behavior of the
model at some special points, and some general properties of the phase diagram. In this section, we
will focus on the case $m=3$. We start by giving a quick overview of the phase diagram in the next
subsection, followed by a more detailed description of the various phases and phase transitions in the
subsequent subsections.

\subsection{Overview of the $m=3$ phase diagram}

We first show the exact diagonalization results of a small system (namely, $L=12$ sites).
\begin{figure}[th]
\begin{center}
\includegraphics[width=\columnwidth]{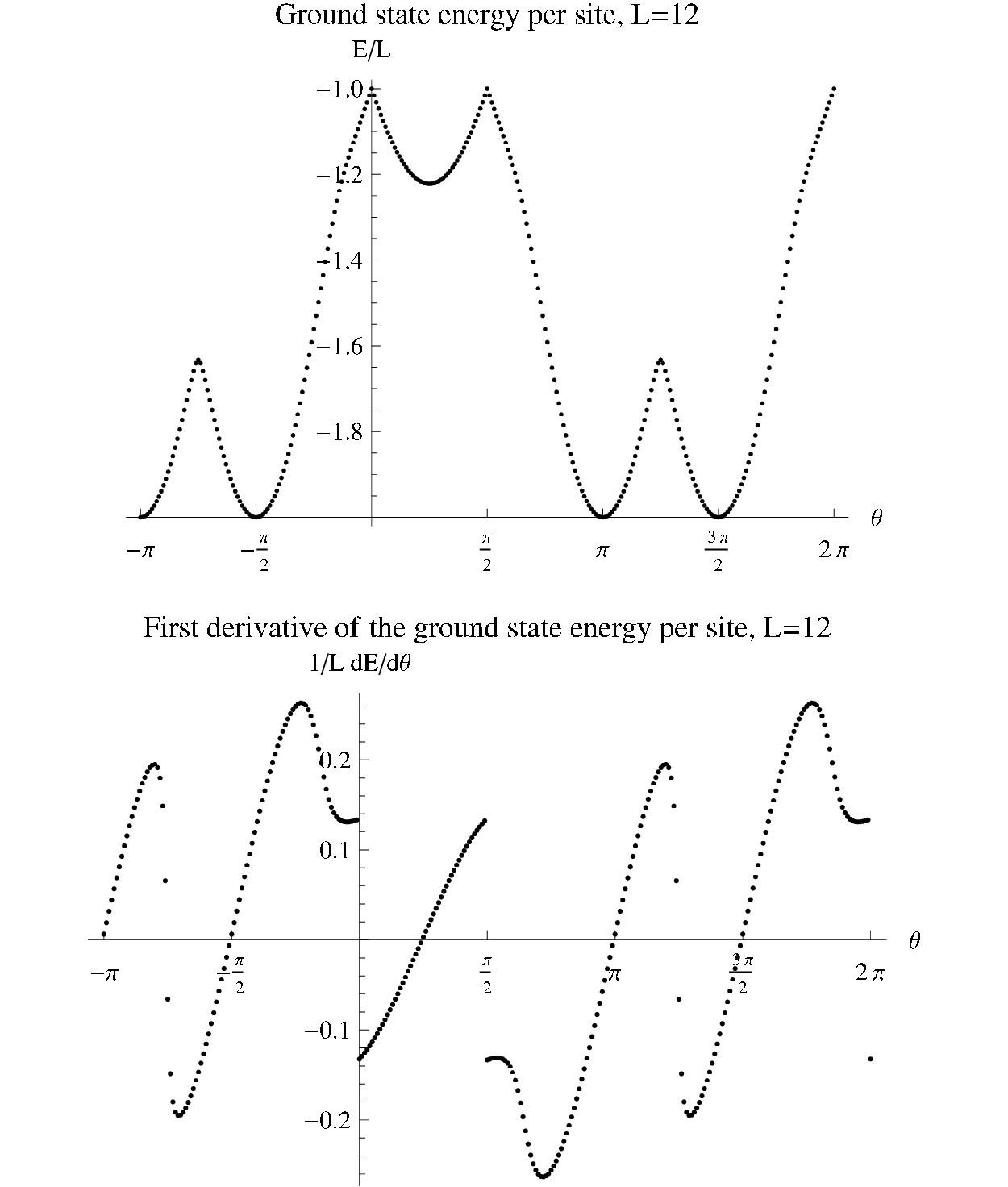}
\caption{Upper panel: ground state energy per site of the $m=3$ model for $L=12$ sites,
as a function of the angle $\theta$. Lower panel: first derivative of the ground state energy
per site of the $m=3$ model for $L=12$ sites, as a function of the angle $\theta$.}
\label{fig:gs-energy-comb}
\end{center}
\end{figure}
In Fig.~\ref{fig:gs-energy-comb}, the ground state energy per site is shown in the upper panel,
while the first derivative is given in the lower panel.

The figures give a clear indication of a first order transition, due to a level crossing, at $\theta=0$. This is signaled by a sharp kink in the ground state energy.
As already discussed, there is at this point also an extensive ground
state degeneracy, which grows exponentially with the system size.
Due to the `duality symmetry',
there is an analogous first order transition at $\theta = \pi/2$.

This first order transition
connects two gapless phases.
In fact, we will later show that throughout the whole region $0< \theta < \pi/2$  the system is
critical and characterized by a central charge $c=2$. This region contains the special point
$\theta = \pi/4$, where the Hamiltonian is self dual. Moreover we will show in the next subsection
{{that} the critical behavior for $\theta\lesssim 0$ can be related to that for $\theta\gtrsim 0$  (and similarly the behaviour for  $\theta\gtrsim\pi/2$ is related to that of $\theta\lesssim\pi/2$) . The critical region for $\theta>\pi/2$ gives
way to a gapped phase at around $\theta \approx 0.6 \pi$, which is signaled by a sudden drop in the first derivative
of the ground state energy, see Fig.\ \ref{fig:gs-energy-comb}.
A much more detailed study of this phase transition will be given in Subsec. \ref{theta0.6piandtheta-0.1pi}.

Beyond $\theta \approx 0.6 \pi$, and in fact in the whole region $0.6\pi \lesssim \theta < 5\pi/4$,
the system is gapped, with a three-fold degenerate ground state in the thermodynamic limit.
In this gapped phase the $\mathcal{U}^y$ symmetry is spontaneously broken. The property of
this phase can be understood by considering the special point $\theta = \pi$, where the
Hamiltonian has no non-commuting operators and can trivially be solved.

The gapped phase in the region $0.6 \pi \lesssim \theta < 5\pi/4$ has its dual analog in the
region $5\pi/4 < \theta \lesssim 1.9 \pi$, where the $\mathcal{U}^x$ symmetry is spontaneously
broken, giving rise to a three-fold degenerate ground state. In addition to the above
the two panels in \Fig{fig:gs-energy-comb} also give hints of
a phase transition between
the two gapped phases at $\theta = 5\pi/4$.
Subsec. \ref{sec:5pi4} is devoted to more detailed discussion
on this phase transition.

\subsection{The behaviour near $\theta = 0$}

Next, let us focus on $\theta\approx 0$, and for simplicity let us concentrate on $m=3$. The
arguments given below can be applied for arbitrary odd $m$.
As discussed above the ground states of the $\cos\theta$ term of the Hamiltonian consists of $|\{q_i\}\rangle$ configurations
where no nearest neighbors have the same $q$. We shall refer to the subspace spanned by these configurations as the projected space. For small $\theta$ the $\sin\theta$ term of the Hamiltonian mixes different configurations within the projected space and also mixes states in the projected space with those outside. However, since there is an energy gap separating the projected space from the rest of the Hilbert space, the effective low-energy Hamiltonian, derived in degenerate perturbation theory to lowest order in $\sin\theta$, is identical to the projection of $H$ on the projective space. The Hamiltonian, with  the ground state energy at $\theta=0$ subtracted, can be written as
\be
H=\sin\theta~{\cal P}~\sum_i P_{i,i+1}~{\cal P}.
\label{th}\ee
where ${\cal P}$ projects states into the projected space, $P_{i,j}$
exchanges
$q_i$ and $q_j$, and the factor $\sin\theta$ defines simply an overall energy scale.

The projected Hamiltonian is symmetric under the same discrete group as the full Hamiltonian. However it has  two additional continuous $\mathrm{U}(1)$ symmetries. These symmetries are generated by the conserved charges $N_q$, which measure the number of sites with the given value of $q$. Due to the constraint $\sum_{q=0}^2 N_q  =L$ only two of the charges are independent.

The spectrum of \Eq{th} is invariant upon a global sign reversal
of the Hamiltonian.
The way to show this is to consider a division of the projected
space into invariant subspaces,
where each of these is spanned by tensor products of $q$-basis vectors,
all having the same numbers $N_{0,1,2}$.
The states within such a subspace can all be mapped into each others by a permutation of the set of $q$ values.
A further subdivision is achieved by collecting all states that are
connected by {\em even} permutations in one group.
This defines two smaller subspaces, with basis vectors that are
interconnected by odd permutations. One can define a unitary operator
$\Gamma_z$ that takes the value $1$ in one of the subspaces
and $-1$ in the other. Since the Hamiltonian (\ref{th}) is
the sum of transposition operators, it only has non-vanishing matrix elements between states with different values for $\Gamma_z$. This means that $H$ anticommutes with $\Gamma_z$,
hence the eigenspectrum of $H$ is symmetric about $E=0$. Upon  $\theta\to-\theta$ the Hamiltonian in \Eq{th} reverses sign. However the eigenspectrum remains unchanged.

Now let us come back to the full Hamiltonian. Near $\theta=0$ the low-lying energy spectrum changes linearly with $\sin\theta$. Therefore, there is a massive crossing of energy levels, with no level repulsion,  at $\theta=0$. In particular the ground state changes abruptly, consistent with a first order transition at $\theta=0$ in the thermodynamic limit ($L\to\infty$).

The above discussion is in reality not restricted to $m=3$, but applies to all odd integer $m$.  Thus, to lowest order in the deviation from $\theta=0$, the low-energy Hamiltonian, for $m =2n+1$, can be written as a projected Hamiltonian of the same form as Eq.\ (\ref{th}).
Similarly one can show that there are now $m-1$ conserved quantities, specified by the
numbers $N_0, \ldots, N_{m-1}$.
The spectrum is inverted in the same way when $\theta\to-\theta$, and consequently, for odd integer $m$, there is at $\theta=0$ a first order phase transition.
Due to the duality symmetry there is precisely the same type of phase transition at point $\theta=\pi/2$.

\subsection{The critical region $0<\theta < \pi/2$}
We now provide evidence for the two phases, one in the interval $0 < \theta < \pi/2$ and the other in the interval $-0.1 \pi\lesssim\theta<0$ ($\pi/2 < \theta\lesssim 0.6\,\pi$), which are connected by the first order phase transition at $\theta=0$ ($\pi/2$), to be gapless.

To determine the properties
of the model in the region
$0<\theta < \pi/2$, we have calculated the entanglement
entropy by DMRG \cite{DMRG_1,DMRG_2}.
In \Fig{entropy} we present a plot of the entanglement entropy $S_A$
for a subsystem (subregion) $A$ of length $\ell=L/2$
for various different values of $\theta$
in the interval $0\le\theta\le\pi/2$.
In \Fig{entropy}(a),
$\theta$ is fixed ($\tan\theta = 1/5$),
while the number $D$ of states kept during the DMRG
calculation is varied. The plots show
that for $L=400$
the value obtained for $D=1000$ has almost converged.
In \Fig{entropy}(b), we compare
the results for various angles $\theta$, keeping $D=1000$ states in each case.

The scaling of the entanglement entropy with system size, shown in \Fig{entropy}, is linear in $\ln L$, which
indicates that for these values of $\theta$ the model
is described by a conformal field theory (CFT).
For (1+1)-dimensional CFTs,
with open boundary condition,
the entanglement entropy $S_A$  of a subsystem (subregion)
$A$ of length $\ell$,
is $S_{A}=(c/6)\log \bigl( (2L/\pi a )\sin (\pi \ell/L) \bigr) +g+c_0$,
\cite{Holzhey1994,Korepin2003,entS}
where
$c$ is the central charge associated with the critical behaviour,
$a$ is a length scale,
$g$ is the boundary entropy,
and $c_0$ a non-universal constant.
In our calculation,we fix $\ell$ to $L/2$, so that $S_{A}\simeq(c/6)\log (L )$ for large $L$.

From the numerical results,
we conclude that for all values of $\theta$ studied
the system exhibits conformal invariance (hence is gapless)
with a central charge $c=2$.
This value of the central charge can be understood by noting that for small $\theta$
there are two conserved charges, each generating an emerging $U(1)$ symmetry.
For a similar gapless phase with $c=2$, realized in interacting boson systems
on the three-leg ladder
at one-third filling, see
Refs.\ \onlinecite{Block2011,Mishmash2011}.
}

\begin{figure}[tbp]
\includegraphics[scale=0.35]{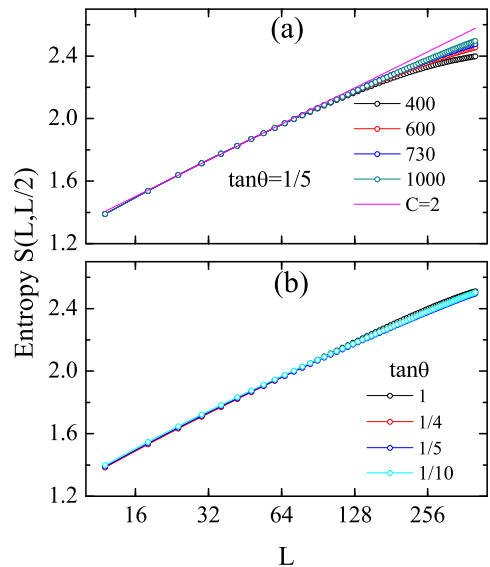}
\caption{
(a) The entanglement entropy associated with $\tan\theta=1/5$ as a function of $\ln L$ under open boundary condition. Different curves are distinguished by the different number of kept states $D$ in the DMRG calculation. The fitting curves are constructed using $c=2$. (b) A comparison of the entanglement entropy for $\tan\theta=
1,1/4,1/5,1/10$ calculated with $D=1000$.
\label{entropy}}
\end{figure}

Entanglement entropy plots for the region $-0.1 \pi\lesssim\theta<0$ and $\pi/2<\theta\lesssim 0.6 \pi$ show the same picture, and here too, we obtain $c\approx 2$ indicating gaplessness (see Fig.5). The fact that the central charge is the same for these two $\theta$ regimes  can be understood as a result of the invariance of the spectrum of \Eq{th} upon reversing the sign of the projected Hamiltonian.
\begin{figure}[tbp]
\includegraphics[scale=0.35]{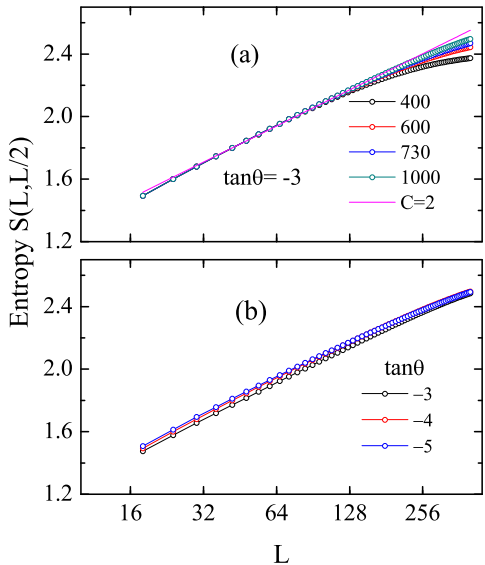}
\caption{
Same types of plots as
Fig.\ \ref{entropy}, here with values of $\theta$ in the interval $\pi/2<\theta\lesssim0.6 \pi$.
\label{entropy2}}
\end{figure}

To give further evidence that the system is critical, we also calculate the entanglement spectrum by means of DMRG. In particular, we consider the eigenvalues $\{\lambda_i\}$of the density matrix. It was shown in \cite{Ent_spec} that the mean number of eigenvalues larger than a given $\lambda$, denoted by $n(\lambda)$ can be calculated from CFT, with the result
\begin{equation}
n(\lambda )=I_{0}(2\sqrt{b\ln (\lambda _{\max }/\lambda )}) \ ,
\end{equation}
where $I_{k}(x)$ is the modified Bessel function of the first kind, $b=-\ln(\lambda _{\max })$ and $\lambda_{\max}$ is the largest eigenvalue of the reduced density matrix.

In Fig.\ \ref{E_spec} we plot the distribution of the $\lambda_i$ for two points in the gapless region. Concretely, we plot the value of the $i$th eigenvalue, following \cite{Ent_spec}. The data agrees with the CFT result asymptotically, which confirms that the gapless phase can be described by CFT.
In principle, it should be possible to extract more information about which CFT describes our system from the distribution of the small eigenvalues of the reduced density matrix, but this is a difficult talks, which we leave for future investigations.

\begin{figure}[tbp]
\includegraphics[scale=0.35]{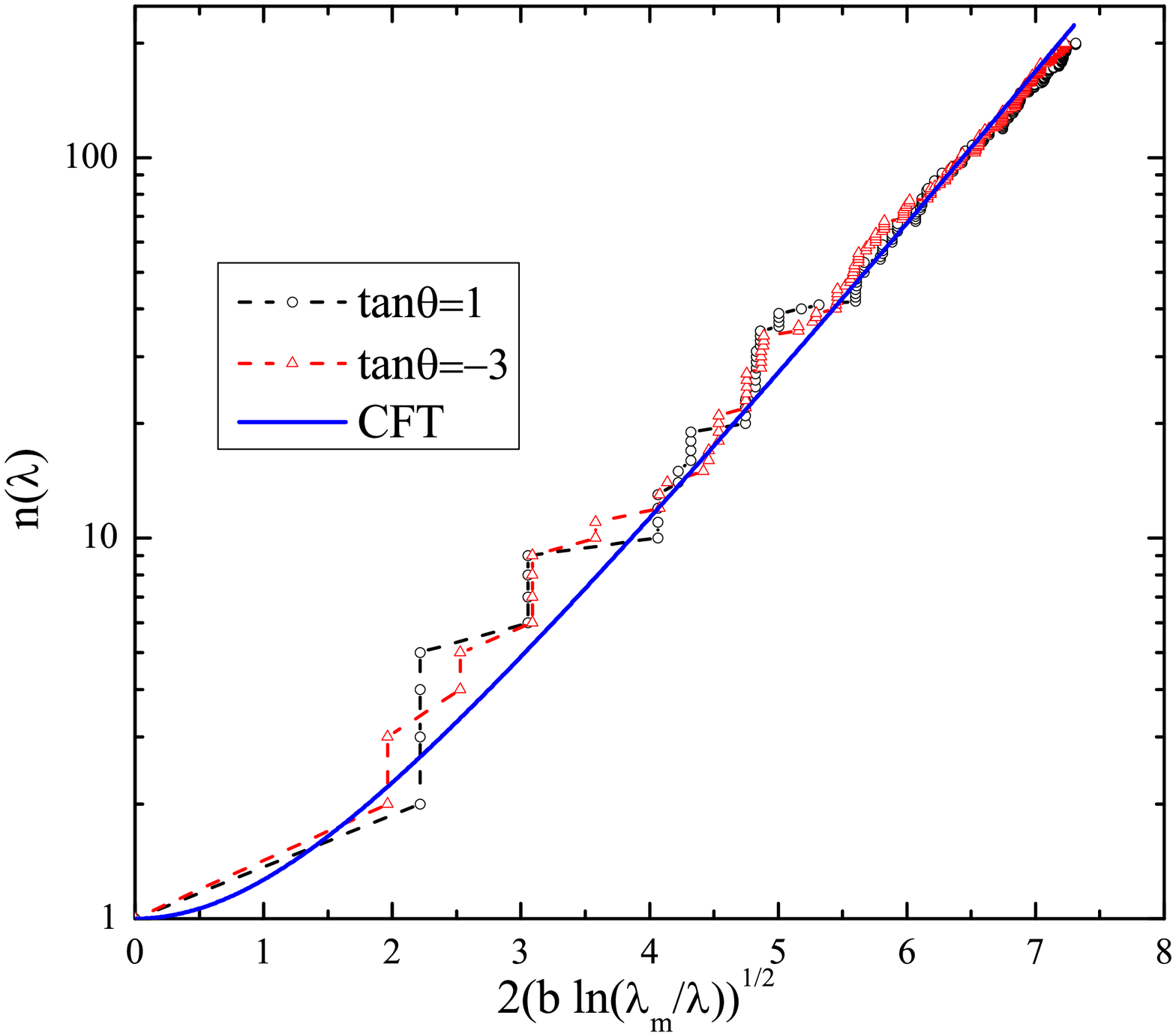}
\caption{
The distribution of the eigenvalues of the density matrix $\lambda$. The length of the chain is 102
and the number of states kept in the DMRG is 1000. The blue solid line is the prediction of CFT. The step structure in the calculated data is due to the degeneracies of the eigenvalue of reduced density matrix.
\label{E_spec}}
\end{figure}

Numerical studies of energy gaps between low lying levels confirm the picture of gapless phases in the intervals referred to above. The DMRG evaluations of the gaps show a clear $1/L$ scaling behavior, consistent with a conformal field theory limit when $L\to\infty$. As we already have pointed out, the numerical plots of the $\theta$ dependence of the ground state energy show a discontinuity in the derivative at the points $\theta=0$ and  $\theta=\pi/2$, consistent with a first order phase transition at these points.

The space of $c=2$ conformal field theories is rather extensive, see Ref.\ \onlinecite{Dulat2000}
for an overview of possible theories.
In the past, in order to identify the low energy conformal field theories of lattice models, it has been very instructive to study the models under periodic boundary conditions, and to calculate the energy spectra as
functions of the various momenta. This can give crucial information in identifying the type of critical
behavior.
\begin{figure}[bh]
\includegraphics[width=1.0\columnwidth]{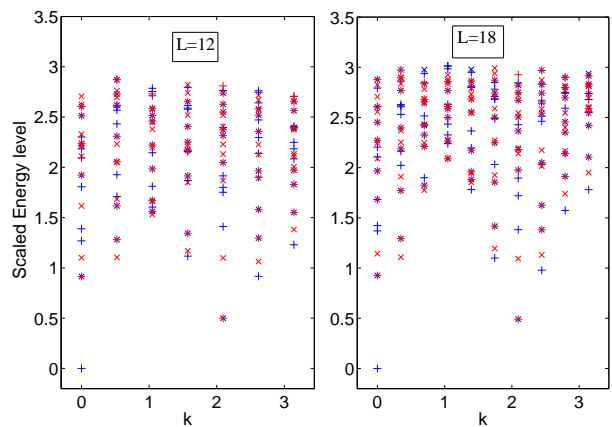}
\caption{
The momentum resolved spectrum at $\theta = \pi/4$, for $L=18$.
The blue plusses correspond to the singlets $(Q,\tilde{Q}) = (0,0)$; the
red crosses to the quadruplets $(Q,\tilde{Q}) = (1,1);(1,2);(2,1);(2,2)$ and the
blue cross, red plus combination to $(Q,\tilde{Q}) = (0,1);(0,2);(1,0);(2,0)$.
}
\label{fig:spectrum-pi4}
\end{figure}
For the torus model we give in  \Fig{fig:spectrum-pi4} the energy spectra, resolved with respect to momentum, for $\theta= \pi/4$, and for system sizes $L=12$ and $L=18$.

From conformal field theory, it follows that if a system is critical, the energies of the low lying states
are given by
\begin{equation}
E = E_{0} L + \frac{2\pi v}{L} \bigl(h_{\rm L} + h_{\rm R} -\frac{c}{12} \bigr) \ .
\end{equation}
In this equation, $E_0$ is a constant energy per site, $v$ is a characteristic velocity, $c$ the
central charge, and $h_{\rm L}$, $h_{\rm R}$ the left and right scaling dimensions of the
associated field in the CFT.
In our case we know from the DMRG results that $c=2$. Furthermore, from studies of the $L$ dependence of the ground state energy,  which has $h_{\rm L} = h_{\rm R} = 0$, the velocity parameter $v$ has been determined.
In Fig.\ref{fig:spectrum-pi4} the energies
are shifted such that the ground state has zero energy, and the energies are rescaled in units such that $2 \pi v/{L} = 1$.

Scaling operators in CFT are grouped in
terms of primary fields, with total scaling dimension $h_{\rm L} + h_{\rm R} = \Delta_p$, and their
descendants, with $h_{\rm L} + h_{\rm R} = \Delta_p + n$, where $n$ is a positive integer.
When the low-energy spectrum is presented in the form shown in \Fig{fig:spectrum-pi4}, one should in principle, be able to identify the CFT through the position $\Delta_p$ of the primary fields, with their associated towers of descendent state. However, in our case we have not been able to make such an identification. In part that is due to the wide range of possibilities for $c=2$, as discussed in Ref.\ \onlinecite{Dulat2000}, but also due to the effects finite size effects for the systems we are able to perform exact diagonalization. Thus, the low energy spectra show several characteristics of a critical system, but we have not been able to conclusively identify which $c=2$ conformal field theory that
describes the critical behavior.

\subsection{The phase transitions at $\theta \approx 0.6\pi$ and $\theta \approx -0.1\pi$}
\label{theta0.6piandtheta-0.1pi}

The numerics indicates that
a phase transition to a gapped phase, with a threefold degenerate ground state, takes place at $\theta\approx -0.1\pi$, with a similar transition at $\theta\approx 0.6 \pi$. These two gapped phases, which are connected by the duality transformation seem to cover the parameter range  $0.6 \pi \lesssim\theta\lesssim 1.9\pi$. We will come back to these gapped
phases in the next subsection.

\Fig{cross}(a) shows the numerical estimate of the location of the critical point between the gapless phase and gapped phase in the vicinity of $\theta\approx 0.6 \pi$.  In the plot the critical point is marked by the crossing between two excitation energies for different values of $L$.
The curve ``gap in the $Q=0$ sector'' is the energy difference between the ground state and the lowest excited state in the $Q=0$ sector, while the curve ``$E_g(Q=1)-E_g(Q=0)$'' gives the gap between the lowest energy state in the $Q=1$ sector and the ground state. In the gapped phase the excited $Q=0$ state will merge with the ground state in the limit $L\to\infty$, but in the gapless phase it lies above the lowest energy state in the $Q=1$ sector. Thus a crossing between the two excited states takes place at a point which moves towards the phase transition point as $L\to \infty$.
The plot shows
a convergence of the crossing towards a point slightly below $\theta=0.61\pi$. A plot of the second derivative of the ground state energy through this point, as shown in \Fig{cross}(b), indicates that a continuous phase transition is taking place.
\begin{figure}[tbp]
\begin{center}
\includegraphics[width= 7.5cm]{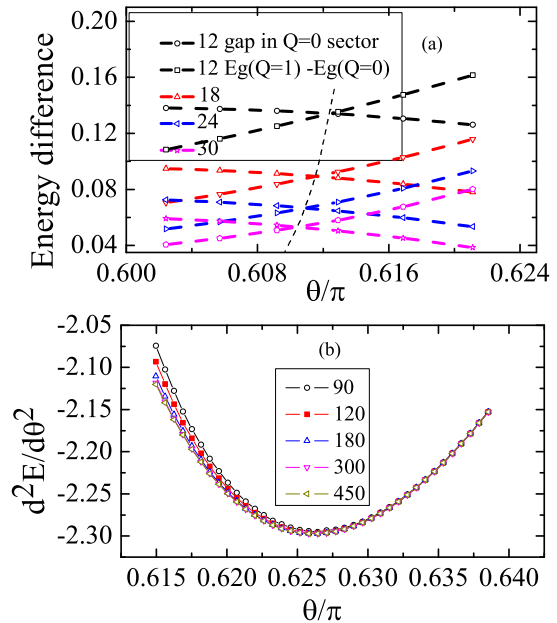}
\caption{
(a)
The crossing between the ground state in $Q=1$
and the first excited state in the $Q=0$ sector.
Different color indicates
different lattice size.
(b) The second derivative of ground state energy with respect to $\theta$.
\label{cross}}
\end{center}
\end{figure}
It is worthy to note the continuous phase transitions in question link a gapless phase and a gapped phase. This is different from usual quantum phase transitions where the phases on both sides are gapped. It is similar to the Kosterlitz-Thouless phase transition except the central charge is different.

\subsection{The gapped phases}

We now focus our attention on the gapped phases of the $m=3$ quantum torus model.
A first gapped phase extends from the phase transition near $\theta \approx 0.6\pi$ to
$\theta = 5\pi/4$. The second gapped phase is dual to the first one, and ranges from
$\theta = 5\pi/4$ to $\theta \approx 1.9\pi$. We will discuss the phase transition between
these two gapped phases in the following subsection.

To understand the symmetry breaking in the gapped phase it is useful to consider the special points $\theta=\pi$ and $\theta=3\pi/2$. For these values of $\theta$  the Hamiltonian becomes classical (i.e., there is no non-commuting operators), and where the ground state is exactly threefold degenerate even for finite $L$.  Away from these points there is finite-size lifting of the ground state degeneracy. However the latter decreases exponentially with increasing $L$.
In  \Fig{gap}(a) the presence of a gap is demonstrated via the saturation of the entanglement entropy as a function of $\ln L$. The exponential decay of the finite-size gap between the states which evolve into the ground state is shown in \Fig{gap}(b).

At the special points $\theta=\pi$ the threefold degenerate ground state is spanned by the product states $\prod_{i=1}^L |q\rangle$, $m=1,2,3$. As a consequence the correlation function $\langle U^{x\dag}_i U^x_j\rangle$ is simply a constant, while the corresponding $U^y$ correlation function vanishes for $i\neq j$. Even if this is a very special situation, a similar behavior of the correlations functions is seen in the full gapped phase, up to $\theta=5\pi/4$. Thus, the $U^x$ correlation function is long range, while the $U^y$ correlation function decays exponentially with the distance between the two points $i$ and $j$. This is shown for a particular value $\theta=1.265\pi$ in \Fig{fig:corr-5pi4} for a system of length $L=102$.
At the point $\theta=5\pi/4$ there is an interchange between the correlations of  $U^x$ and $U^y$, as follows from the duality symmetry.  The curve corresponding to $\theta=5\pi/4$ in \Fig{fig:corr-5pi4} shows identical long range correlations for $U^x$ and $U^y$. This can be understood as due to the fact that the ground state for any finite $L$ is a 50/50 superposition of two states, with correlations that are symmetric under the interchange $x\leftrightarrow y$.

\begin{figure}[tbp]
\includegraphics[scale=0.35]{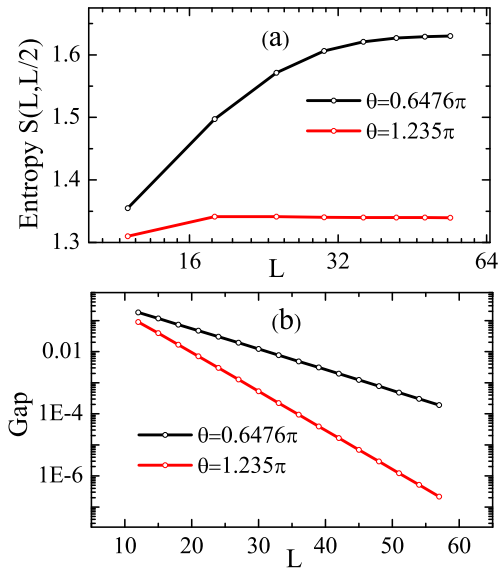}
\caption{
(a)
The entanglement entropy for two different $\theta$ in the gapped region.
The entropy saturates as $\ln L$.
(b) The energy difference of the first excited state and the ground state
in the $Q=0$ sector. The finite size gap decays exponentially to $0$
with increasing $L$.
}
\label{gap}
\end{figure}
\begin{figure}[bh]
\includegraphics[width=.8\columnwidth]{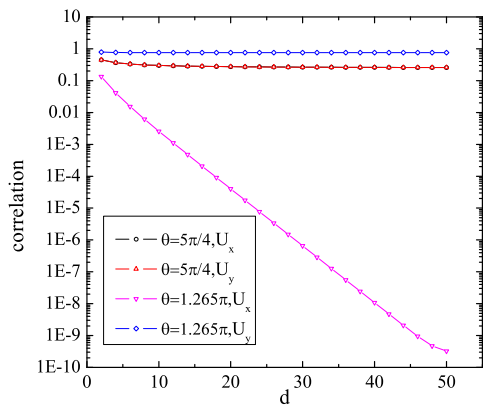}
\caption{
Logarithmic plot of the ${U}^x$ and ${U}^y$ correlations
at $\theta = 5\pi/4$
and $\theta = 1.265\pi$. For $\theta=5\pi/4$ the two curves fall on the top of each other.
$d$ is the distance between the two points for which the correlation is calculated.
The system size used in the calculation is $L=102$.}
\label{fig:corr-5pi4}
\end{figure}

\subsection{The phase transition at $\theta = 5\pi/4$}
\label{sec:5pi4}

As a final point,
we have examined the phase transition at the point $\theta=5\pi/4$,
at which the two gapped phases meet.
In the vicinity of $\theta=5\pi/4$ there are six energy eigenstates that are important: two $(Q,\tilde{Q}) = (0,0)$ singlets, a $(Q,\tilde{Q}) = (0,1);(0,2)$ doublet and another  $(Q,\tilde{Q}) = (1,0);(2,0)$
doublet. All the numerical evidences are consistent with the following picture in the thermodynamic limit. For $\theta < 5\pi/4$ the ground state is triply degenerate. The triplet involves one of the  $(Q,\tilde{Q}) = (0,0)$ singlet and the $(Q,\tilde{Q}) = (0,1);(0,2)$ doublet. For $\theta > 5\pi/4$ the ground state is also triply degenerate. This time the triplet involves the other $(Q,\tilde{Q}) = (0,0)$ singlet and the $(Q,\tilde{Q}) = (1,0);(2,0)$ doublet. At $\theta=5\pi/4$ the two triplets cross, resulting in a six fold degenerate ground state. For finite $L$ the doublet $(Q,\tilde{Q}) = (0,1);(0,2)$ precisely degenerate with $(Q,\tilde{Q}) = (1,0);(2,0)$. This is because they form the four dimensional irreducible representation of the enlarged symmetry group at the selfdual point. In contrast the two $(Q,\tilde{Q}) = (0,0)$ singlets are slightly split due to avoided crossing caused by the finite system size.
The above picture suggests a first order phase transition at the selfdual point in the thermodynamic limit. This is caused by the crossing of energy levels. The two phases are  distinguished by different long range correlations for either $U^x$ or $U^y$, as previously discussed.

The character of the phase transition is further illustrated by the plots in
\Figs{gap_a-1b-1}, \ref{a-1b-1} and \ref{entropy-compare}.
In \Fig{gap_a-1b-1}, we display the behavior of the gap
at the transition point $\theta = 5\pi/4$. The ground state is a $(Q,\tilde{Q}) = (0,0)$
singlet, and $\Delta_1$ denotes the energy difference with the first excited `level', which
is the $(Q,\tilde{Q}) = (0,1);(0,2);(1,0);(1,0)$ quadruplet (see the discussion above).
$\Delta_2$ denotes the energy difference between the ground state and the second
excited `level', again a $(Q,\tilde{Q}) = (0,0)$ singlet. The energy difference $\Delta_1$ and
$\Delta_2$ are given in \Fig{gap_a-1b-1} for various values of $D$, the number of
states kept in the DMRG, clearly showing that the results have converged as a function
of $D$. Both the energies of the first and second excited levels decay exponentially
with system size, which implies a six-fold degenerate ground state, because other levels
have a finite gap to these six degenerate states in the thermodynamic limit, see also
\Fig{a-1b-1}.

Panel $(a)$ of \Fig{a-1b-1} shows the behavior of the energy difference between the
two lowest lying levels in the $Q=1$ sector. The latter does not merge with the degenerate ground state in the thermodynamic limit.
The plot shows that a small energy gap remains as $L\to\infty$.

In panel (b),
the first derivative of the ground state energy is plotted as a function of $\theta$. It shows
a sharp drop at $\theta = 5\pi/4$, which is expected to become infinitely sharp in the
limit $L\rightarrow\infty$, because transition happens at an isolated point between two gapped
phases. The sharp change in the derivative of the ground state energy is consistent with the picture of the phase transition as caused by the crossing of the two lowest energy levels.

\begin{figure}[tbp]
\begin{center}
\includegraphics[scale= 0.3]{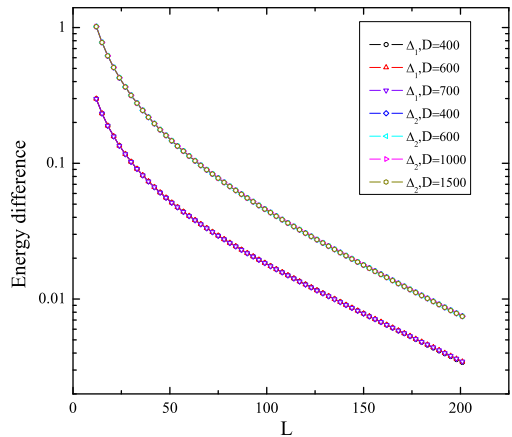}
\caption{
The exponential decay of the energy difference between
the two lowest levels and the ground state at $\theta=5\pi/4$.
$\Delta_1 $ is the
energy difference between the first excitation level (a quadruplet, see the main text)
and the ground state (a singlet), and
$\Delta_2 $ is energy difference between the second excitation level (also a singlet)
and  the ground state.
They both decay to zero exponentially with system size $L$.
\label{gap_a-1b-1}}
\end{center}
\end{figure}

\begin{figure}[tbp]
\begin{center}
\includegraphics[scale= 0.3]{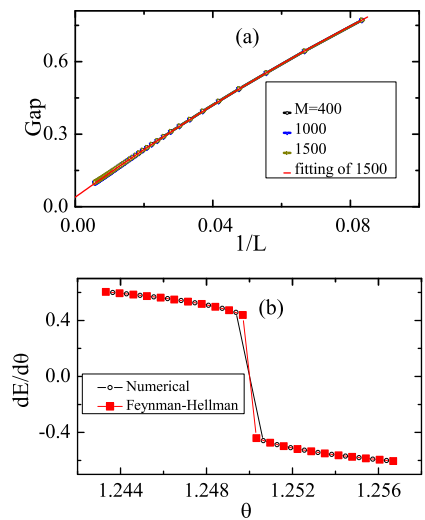}
\caption{
(a)
The energy difference between the first excited state
and
the ground state in the $Q=1$ sector
at $\theta=5\pi/4$.
We use
a quadratic fitting of the gap with $1/L$ using the $D=1500$ data.
As $L$ goes to infinity, there is a small gap $(\sim 0.04)$.
(b)
The first derivative of ground state energy with respect to $\theta$.
The circle result are got by the numerical differential
of ground state energy
while the square data are got directly by using the Feynman-Hellman theorem.
\label{a-1b-1}}
\end{center}
\end{figure}
\begin{figure}[tbp]
\begin{center}

\includegraphics[width=0.9 \columnwidth]{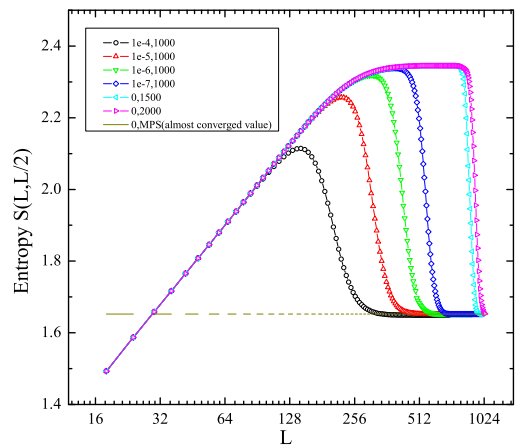}
\caption{
The entanglement entropy,
close to the point $\theta=5\pi/4$,
as obtained
from DMRG and the MPS method.}
\label{entropy-compare}
\end{center}
\end{figure}

In \Fig{entropy-compare},
we show the results of a detailed study of the behavior
of the entanglement entropy, near the point $\theta = 5\pi/4$.
The plot shows that for system size up to about $L=256$,
the entanglement entropy behaves as expected for a critical point, and is consistent
with $c=2$. For larger system sizes, the entanglement entropy flattens of, and finally crosses
over to the (lower) value obtained via the MPS method.
This last crossing over happens
at larger system size, if the number of states kept in the DMRG calculation is increased,
although for $D=2000$, the numerics seems to have converged.

The drop in the entanglement entropy for large $L$, at $\theta \sim 5\pi/4$,
can be understood in terms of the level-crossing picture as follows:
There are two states with $(Q,\tilde{Q})=(0,0)$ participating in the crossing;
let us denote them
$|\psi_-\rangle$
and
$|\psi_+\rangle$,
where
$|\psi_-\rangle$ ($|\psi_+\rangle$) is a ground state for
$\theta<5 \pi/4$ ($\theta>5 \pi/4$).
They are not orthogonal but the overlap is exponentially suppressed for large $L$.
For finite $L$,
the true ground state is a mixture of $|\psi_-\rangle$ and $|\psi_+\rangle$,
which at $\theta=5 \pi/4$ is
an equal-superposition thereof.
Away from this point the ground state rapidly rotates into either
$|\psi_-\rangle$
or
$|\psi_+\rangle$,
depending on whether $\theta$ is reduced or increased from $\theta=5\pi/4$.
This rotation is more rapid the larger $L$
is so that for $L\to \infty$ the crossing between $|\psi_-\rangle$ and $|\psi_+\rangle$
becomes sharp.
Assume we choose $\theta$ slightly smaller than $5\pi/4$.
For small $L$ the ground state is essentially the 50
percent mixed state, but when $L$ increases at some point it rapidly changes to $|\psi_-\rangle$.
The entanglement entropy of each of the two states $|\psi_+\rangle$ and $|\psi_-\rangle$ clearly is smaller than that of the superposition of the two. Due to the symmetry between $|\psi_-\rangle$ and $|\psi_+\rangle$ at $\theta=5\pi/4$,
and the rapid decay of the overlap between them as $L\to\infty$, we can in fact estimate the
drop in entanglement entropy. Since each of the two states in the superposition gives equal contributions to the entanglement entropy the drop should be close to $\ln 2$. The numerical value found for the curves plotted in \Fig{entropy-compare} is in fact very close to $\ln 2$.

We have conclusively shown that the gap at the transition point at $\theta = 5\pi/4$
does not close, and that that system exhibits long-range order. It is intriguing to observe
that the scaling of the entanglement entropy shows behavior consistent with $c=2$
critical behavior) up to fairly large system sizes of at least $L=200$ (due to the rather
large correlation length). Combined with exact diagonalization results (which we
performed up to $L=18$, not shown), one could incorrectly be led to believe that the
system is critical. It would be interesting to investigate if the system can be (fine-) tuned
to become critical, by perturbing away from the gapped $\theta= 5\pi/4$ point.
One could for instance allow complex amplitudes in the Hamiltonian \eqref{tr}, or
mixed terms such as $U^{x}_{i} U^{y}_{i+1}$.

\section{Conclusions}

To summarize, we have presented a class of one-dimensional lattice models, called the quantum torus chain. These models  depend on an integer-valued parameter $m$ (the number of magnetic flux quanta piercing through each torus) which controls the presence/absence of frustration. In addition there is a continuous parameter $\theta$ which controls the ratio of the non-commuting terms in the Hamiltonian. The models show a variety of gapless and gapped phases. In all cases the gapped phases break symmetries of the Hamiltonian. We characterize these gapped phases based on the properties of specific points  in parameter space, where the Hamiltonian is exactly solvable. For the gapless phases, we have used DMRG and matrix product state methods to evaluate the entanglement entropy and hence deduce the central charge.

Based on these analysis we have concluded that while for even $m$ the system is generically gapped with isolated quantum critical points separating different gapped phases. For odd $m$ there exists extended regions in the parameter $\theta$ where the system is gapless. The above general conclusion is motivated by the mapping of the $m=2$ model on the spin 1/2 XZ model, and the detailed numerical studies of the $m=3$ model. For the latter the quantum phase transitions between two gapless or two gapped phases are first order, while the transitions between gapless and gapped phase are continuous. Consistent with these conclusions we have numerical evidence from studies of the $m=4$ model that this is also gapped in the first quadrant $0<\theta<\pi/2$.

Our study raises a series of open questions.
At the moment we have not been able to pin down the conformal field theory for the $c=2$ gapless phases of the $m=3$ model. Of course gapless phases have precedents in one dimension (e.g. the Luttinger liquid\cite{lut}). However gapless {\em phases} in more than one dimension are very unusual. Our model of course can be defined in any spatial dimension.

Without going into much detail, we can say that for $m$ even, there is no frustration and at the special angles $\theta = 0, \pi/2, \pi, 3\pi/2$, the ground states are direct product states, with a finite degeneracy. Thus at least near these points, we expect the system to be gapped. Whether or not these gapped phases extend over the full parameter space, leaving only critical points as is the case in our 1-d model needs further investigation. For odd $m$, there is also quantum frustration in higher dimensions. It is interesting to ask whether the gapless phase still exists in higher dimensions.

There are several ways to extend our model. One could try to include terms which mimic a magnetic field in ordinary spin chains. In our model, the closest analog of a magnetic field would be a one-site term, which breaks the symmetry of the model, by favoring (say) one of the possible states. More phase transitions should be expected in this enlarged parameter space, because a strong symmetry breaking term will simply `polarize the degrees of freedom', leading to ordered phases.

The general properties we find in the this work are consistent with the results discussed by Chen {\it et al.} \cite{Chen11} for systems where the internal symmetry group is represented locally on each site by a projective representation. It will be interesting to consider generalizations of the torus model where the symmetry is represented linearly locally. In this case symmetry protected topological states become possible. Finally it is interesting to ask whether it is possible to describe the gapless phase in terms of a discrete non-linear $\sigma$ model with a topological term.
We will leave these and other related questions as interesting projects for future investigation. \\

{\it Acknowledgement:}
DHL is supported by DOE grant number DE-AC02-05CH11231.

\end{document}